\documentclass[sigconf]{acmart}

\AtBeginDocument{%
  \providecommand\BibTeX{{%
    \normalfont B\kern-0.5em{\scshape i\kern-0.25em b}\kern-0.8em\TeX}}}

\usepackage{algorithm}
\usepackage{algpseudocode}
\usepackage{multirow}
\usepackage{arydshln}
\usepackage{amstext}
\usepackage{enumitem}
\newcommand{\tabincell}[ 2 ]{\begin{tabular}{@{}#1@{}}#2\end{tabular}}
\usepackage{amsmath}
\usepackage{bm}

\makeatletter
\g@addto@macro\normalsize{%
  \abovedisplayskip 4.2pt plus1pt 
  \belowdisplayskip
  \abovedisplayskip
  \abovedisplayshortskip  4pt plus1pt%
  \belowdisplayshortskip  4pt plus1pt
}
\makeatother

\copyrightyear{2021}
\acmYear{2021}
\setcopyright{acmcopyright}\acmConference[WSDM '21]{Proceedings of the Fourteenth ACM International Conference on Web Search and Data Mining}{March 8--12, 2021}{Virtual Event, Israel}
\acmBooktitle{Proceedings of the Fourteenth ACM International Conference on Web Search and Data Mining (WSDM '21), March 8--12, 2021, Virtual Event, Israel}
\acmPrice{15.00}
\acmDOI{10.1145/3437963.3441777}
\acmISBN{978-1-4503-8297-7/21/03}
\begin{document}
\fancyhead{}
\title{PROP: Pre-training with Representative Words Prediction for Ad-hoc Retrieval}
\author{Xinyu Ma, Jiafeng Guo, Ruqing Zhang, Yixing Fan, Xiang Ji and Xueqi Cheng}
\affiliation{
  \institution{
    CAS Key Lab of Network Data Science and Technology, Institute of Computing Technology, \\ Chinese Academy of Sciences, Beijing, China\\
    University of Chinese Academy of Sciences, Beijing, China\\} 
}
\email{{maxinyu17g,guojiafeng,zhangruqing,fanyixing,jixiang19s,cxq}@ict.ac.cn}

\renewcommand{\shortauthors}{Ma and Guo, et al.}

\begin{abstract}
Recently pre-trained language representation models such as BERT have shown great success when fine-tuned on downstream tasks including information retrieval (IR). However, pre-training objectives tailored for ad-hoc retrieval have not been well explored. In this paper, we propose Pre-training with Representative wOrds Prediction (PROP) for ad-hoc retrieval. PROP is inspired by the classical statistical language model for IR, specifically the query likelihood model, which assumes that the query is generated as the piece of text representative of the ``ideal'' document. Based on this idea, we construct the representative words prediction (ROP) task for pre-training. Given an input document, we sample a pair of word sets according to the document language model, where the set with higher likelihood is deemed as more representative of the document. We then pre-train the Transformer model to predict the pairwise preference between the two word sets, jointly with the Masked Language Model (MLM) objective. By further fine-tuning on a variety of representative downstream ad-hoc retrieval tasks, PROP achieves significant improvements over baselines without pre-training or with other pre-training methods. We also show that PROP can achieve exciting performance under both the zero- and low-resource IR settings.
The code is available at \url{https://github.com/Albert-Ma/PROP}.

\end{abstract}



\maketitle

\section{Introduction}\label{sec:intro}

Recent advances have shown that pre-trained language representation models, such as OpenAI GPT~\cite{radford2018improving}, BERT~\cite{devlin2018bert} and XLNET~\cite{yang2019xlnet}, can capture rich language information from text, and achieve state-of-the-art accuracy in many downstream natural language processing (NLP) tasks including summarization~\cite{sutskever2014seq2seq}, sentiment classification~\cite{socher2013recursive}, and named entity recognition~\cite{sang2003introduction}, which usually have limited supervised data. The success of pre-trained models in NLP has also attracted a lot of attention in the IR community. Researchers have explored the popular models, e.g., ELMo \cite{peters2018deep} and BERT, in the context of ad-hoc document ranking, and showed that they can also largely benefit the search tasks where training data are limited~\cite{nogueira2019passage,nogueira2019multi,dai2019deeper,yang2019simple,macavaney2019cedr}. 


Despite the exciting performance of pre-trained models on IR tasks, however, pre-training objectives tailored for ad-hoc retrieval have not been well explored. 
On the one hand, most existing pre-training objectives that come from NLP can be summarized into two folds, i.e., sequence-based and sentence pair-based tasks. 
Sequence-based pre-training tasks, such as Masked Language Modeling (MLM) \cite{devlin2018bert} and Permuted Language Modeling (PLM)~\cite{yang2019xlnet}, aim to learn contextual representations for a word based on the sequence-level co-occurrence information.
Sentence pair-based pre-training tasks, such as Next Sentence Prediction (NSP)~\cite{devlin2018bert} and Sentence Order Prediction (SOP)~\cite{wang2019structbert}, attempt to teach the model to better understand the inter-sentence coherence relationship~\cite{qiu2020pre}.
On the other hand, IR tasks such as ad-hoc retrieval typically handle short (keyword-based) queries and long (multi-sentence-based) documents. It requires not only understanding the text content of a query and a document, but also modeling the relevance relationship between the two.
When we look at those existing pre-training objectives from the IR perspective, we may find that: 1) Sequence-based pre-training tasks could in general contribute to build good contextual representations for the query and the document; 2) The learning objectives of sentence pair-based tasks, however, are quite diverged from the IR requirement, not just due to the input difference (sentence-pair vs. query-document) but also the relation type (coherence vs. relevance).
It is generally hypothesized that using a pre-training objective that more closely resembles the downstream task leads to better fine-tuning performance \cite{zhang2019pegasus}.  In this sense, we argue that the power of pre-training has not been fully exploited for ad-hoc retrieval tasks.

Yet there has been little effort to design pre-training objectives towards ad-hoc retrieval. The most related work in this direction focused on passage retrieval in question answering (QA)~\cite{lee2019latent,chang2020pre}, where three types of pre-training tasks have been proposed including: (1) Inverse Cloze Task (ICT): The query is a sentence randomly drawn from the passage and the document is the rest of sentences; (2) Body First Selection (BFS): The query is a random sentence in the first section of a Wikipedia page, and the document is a random passage from the same page; and (3) Wiki Link Prediction (WLP): The query is a random sentence in the first section of a Wikipedia page, and the document is a passage from another page where there is a hyperlink link to the page of the query. As we can see, these tasks attempt to resemble the relevance relationship between natural language questions and answer passages. Some tasks even depend on certain special document structure, e.g., hyperlink. When applying pre-trained models based on these tasks to ad-hoc retrieval, marginal benefit could be observed on typical benchmark datasets as shown in Section \ref{exp}.



In this paper, therefore, we aim to design a novel pre-training objective tailored for IR which more closely resembles the relevance relationship between query and document in ad-hoc retrieval. The key idea is inspired by the traditional statistical language model for IR, specifically the query likelihood model \cite{ponte1998language} which was proposed in the last century. The query likelihood model assumes that the query is generated as the piece of text representative of the ``ideal'' document \cite{liu2005statistical}. Based on the Bayesian theorem, the relevance relationship between query and document could then be approximated by the query likelihood given the document language model under some mild prior assumption. Based on the classical IR theory, we propose the Representative wOrds Prediction (ROP) task for pre-training. Specifically, given an input document, we sample a pair of word sets according to the document language model, which is defined by a popular multinomial unigram language model with Dirichlet prior smoothing. The word set with higher likelihood is deemed as more ``representative'' of the document. We then pre-train the Transformer model to predict the pairwise preference between the two sets of words, jointly with the Masked Language Model (MLM) objective. The pre-trained model, namely PROP for short, could then be fine-tuned on a variety of downstream ad-hoc retrieval tasks. The key advantage of PROP lies in that it roots in a good theoretical foundation of IR and could be universally trained over large scale text corpus without any special document structure (e.g. hyperlinks) requirement.

We pre-train PROP based on two kinds of large scale text corpus respectively. One is the English Wikipedia which contains tens of millions of well-formed wiki-articles, and the other is the MS MARCO Document Ranking dataset which contains about 4 million Web documents. We then fine-tune PROP on 5 representative downstream ad-hoc retrieval datasets, including Robust04, ClueWeb09-B, Gov2, MQ2007 and MQ2008. The empirical experimental results demonstrated that PROP can achieve significant improvements over baselines without pre-training or with other pre-training methods, and further push forward the state-of-the-art.
Large-scale labeled IR datasets are rare and in practice it is often time-consuming to collect sufficient relevance labels over queries. The most common setting is that of zero- or low-resource ad-hoc retrieval. We simulate both settings and show that our model is capable of obtaining state-of-the-art results when fine-tuning with small numbers of supervised pairs. 
The contributions of this work are listed as follows:
\begin{itemize}[leftmargin=*]
\item We propose PROP, a new pre-training objective for ad-hoc retrieval which has a good theoretical IR foundation and could be universally trained over large scale text corpus without any special document structure requirement. 
\item We evaluate PROP on a variety of downstream ad-hoc retrieval tasks and demonstrate that our model can surpass the state-of-the-art methods. 
\item We show how good ad-hoc retrieval performance can be achieved across different datasets with very little supervision by fine-tuning the PROP model.
\end{itemize}

\section{BACKGROUND}\label{sec:backg}
We first briefly review the classical statistical language model for IR, specifically the query likelihood model, which is the theoretical foundation of our pre-training method.
The basic idea of the query likelihood model assumes that the user has a reasonable idea of the terms that are likely to appear in the ``ideal'' document that can satisfy his/her information need \cite{ponte1998language}.

The query is thus generated as the piece of text representative of the ``ideal'' document~\cite{liu2005statistical}. 

Such a query-generation idea could then be formulated as a probabilistic model using the Bayesian theorem.
Specifically, given a query $Q=q_{1}...q_{m}$ and a document $D=w_{1}...w_{n}$, we have: 
\begin{equation}
 P(D|Q) \propto P(Q|\theta_D)P(D),
\end{equation}
where $\theta_D$ is a document language model estimated for every document.
The prior probability $P(D)$ is usually assumed to be uniform across all documents and thus can be ignored. Based on this simplification, the estimation of the relevance of a document to a query $P(D|Q)$ could be approximated by the query likelihood $P(Q|\theta_D)$, i.e., the query generation probability, according to the document language model $\theta_D$.


Different methods have been proposed for the document language model $\theta_D$, among which a multinomial unigram language model has been most popular and most successful \cite{zhai2008statistical}. Assuming a multinomial language model, one would generate a sequence of words by generating each word independently. In this way, the query likelihood would be 
\begin{equation}\label{eq:unigramLM}
\begin{aligned}
 P(Q|D) &= \prod_{i}^{m} P(q_{i}|\theta_D)\\
 &=\prod_{w\in V} P(w|\theta_D)^{c(w,Q)},
\end{aligned}
\end{equation}
where $V$ is the corpus vocabulary and $c(w,Q)$ is the count of word $w$ in query $Q$.

To better estimate the document language model and eliminate zero probabilities for unseen words, many smoothing methods have been proposed to improve the accuracy of the estimated language model. Among all these methods, Dirichlet prior smoothing appears to work the best, especially for keyword queries (non-verbose queries)~\cite{zhai2017study}, which is defined as
\begin{equation}\label{eq:dirichlet}
P(w|D) = \frac{c(w,D)+\mu P(w|C)}{|D|+\mu},
\end{equation}
where $c(w,D)$ is the count of word $w$ in document $D$, $|D|$ is the length of document $D$ (i.e., the total word counts), $P(w|C)$ is a background (collection) language model estimated based on word counts in the entire collection and $\mu$ is a smoothing parameter. For more details about statistical language models for IR, we refer reader to these papers~\cite{liu2005statistical,zhai2008statistical,manning2010introduction}. 



\begin{algorithm}[t]
\caption{Sampling a Pair of Representative Word Sets}
\label{alg:sampling}
\begin{algorithmic}[1]
\State \textbf{Input:} Document $D$, Vocabulary $V=\{w_{i}\}_{1}^N$, probability of word $w_i$ generated by the document language model with Dirichlet smoothing $P(w_i|D)$, Query likelihood score function $QL(w_i,D)$
\State // {\itshape Choose length}
\State $l = Sample(X)$, $x\sim{Poisson(\lambda)}, x=1,2,3...$ 
\State $S_1,S_2 = \emptyset,\emptyset $
\State // {\itshape Paired Sampling}
\For{$k \gets 1$ to $l$}
    \State $S_1 = S_1 \cup Sample(V), w_i \sim P(w_i|D)$
    \State $S_2 = S_2 \cup Sample(V), w_i \sim P(w_i|D)$
\EndFor
\State // {\itshape Higher likelihood deemed as more representative}
\State $S_1\_score =\prod_{i}^{l} QL(w_i,D), w_i \in S_1$ 
\State $S_2\_score =\prod_{i}^{l} QL(w_i,D), w_i \in S_2$
\If{$S_1\_score>S_2\_score$}
\State \textbf{Output:}($S_1^+,S_2^-,D$)
\Else
\State \textbf{Output:}($S_1^-,S_2^+,D$)
\EndIf
\end{algorithmic}
\end{algorithm}

\section{PROP}\label{sec:prop}
In this section, we present the new pre-training objective PROP which is tailored for ad-hoc retrieval in detail. We also provide some discussions on the differences and connections of PROP with respect to weak supervision methods for IR and existing pre-training objectives respectively.

\subsection{Pre-training Methods}\label{sec:pre-method}
Existing work has demonstrated that using a pre-training objective that more closely resembles the downstream task leads to better fine-tuning performance \cite{song2019mass,zhang2019pegasus}. Given our intended use for ad-hoc retrieval, we aim to introduce a new pre-training task that better resembles the relevance relationship between query and document in IR.

The key idea is inspired by the above query likelihood model which assumes that the query is generated as the piece of text representative of the “ideal” document. 
Based on this assumption, we construct the representative words prediction (ROP) task for pre-training. Specifically, given an input document, we sample a pair of word sets according to the document language model. Intuitively, each sampled word set could be viewed as a generated \text{pseudo query} from the document. In this way, the word set with higher likelihood is deemed as a more ``representative'' query of the document. We then pre-train the Transformer model to predict the pairwise preference between the two word set, i.e., the ROP objective, jointly with Masked Language Model (MLM) objective. The pre-trained model is named as PROP for short. The detailed pre-training procedures are as follows.

\textbf{Representative Word Sets Sampling.~} Given a document, we sample a pair of word sets, each as a generated pseudo query, according to the document language model. To simulate the varied query length in practice \cite{azzopardi2007building,arampatzis2008study}, we first use a Poisson distribution~\cite{consul1973generalization} to sample a positive integer $l$ as the size of the word set, which is defined by
\begin{displaymath}
  P(x) = \frac{\lambda^x e^{-\lambda}}{x},x=1,2...,
\end{displaymath}
where $\lambda$ is a hyper-parameter that indicates the expectation of interval. We then sample a pair of word sets $S_1$ and $S_2$ with the same size $l$ in parallel according to the document language model. Specifically, for each word set, $l$ words are sampled from the corpus vocabulary $V=\{w_{i}\}_{1}^N$ independently according to the multinomial unigram language model with Dirichlet prior smoothing as defined by Equation~(\ref{eq:dirichlet}). The detailed sampling process is shown in Algorithm~\ref{alg:sampling}.

\textbf{Representative Words Prediction (ROP).~} 
Given the pair of word sets sampled above, we compute the likelihood of each set according to Equation~(\ref{eq:unigramLM}), and the set with the higher likelihood is regarded as more representative for the document. We then pre-train a Transformer model to predict the pairwise preference between the two word sets. 

Specially, the word set $S$ and the document $D$ are concatenated as a single input sequence and fed into the Transformer with special delimiting tokens, i.e., $[CLS] + S + [SEP] + D + [SEP]$. 
Each word in the concatenated sequence is represented by summing its distributed, segment, and positional embeddings. 
Then, the hidden state of the special token [CLS], i.e. $\mathbf{H}^{[CLS]}$, is obtained by, 
\begin{equation}
\mathbf{H}^{[CLS]} = Transformer_{L}([CLS] + S + [SEP] + D + [SEP]),
\end{equation}
where $L$ is a hyper-parameter denoting the number of Transformer layers. 
Finally, the likelihood $P(S|D)$, which denotes how representative the word set is to the document, is obtained by applying a multi-layer perceptron (MLP) function over the  $\mathbf{H}^{[CLS]}$ following previous studies~\cite{nogueira2019passage,dai2019deeper,nogueira2019multi}.

Now we denote the pair of word sets sampled and the corresponding document as a triple $(S_1, S_2, D)$. Suppose set $S_1$ has a higher likelihood score than $S_2$ according to Equation~(\ref{eq:unigramLM}), the ROP task can then be formulated by a typical pairwise loss, i.e., hinge loss, for the pre-training.
\begin{equation}\label{irloss}
\begin{aligned}
\mathcal{L}_{ROP} = max(0, 1-P(S_1|D)+s(S_2|D)),
\end{aligned}
\end{equation}

\textbf{Masked Language Modeling (MLM).~} MLM is firstly proposed by Taylor~\cite{taylor1953cloze} in the literature, which is a fill-in-the-blank task. 
MLM first masks out some tokens from the input and then trains the model to predict the masked tokens by the rest of the tokens. As mentioned in the Introduction, the MLM objective could in general contribute to building good contextual representations for the query and the document. Therefore, similar to BERT, PROP also adopts MLM as one of its pre-training objectives besides the pairwise preference prediction objective.

Specifically, the MLM loss $\mathcal{L}_{MLM}$ is defined as: 
\begin{equation}\label{mlmloss}
\begin{aligned}
\mathcal{L}_{MLM} = -\sum_{\hat{x}\in m(\textbf{x})} \log p(\hat{x}|\textbf{x}_{\backslash m(\textbf{x})}),
\end{aligned}
\end{equation}
where $\textbf{x}$ denotes the input sentences, $m(x)$ and $\textbf{x}_{\backslash m(\textbf{x})}$ denotes the masked words and the rest words from $\textbf{x}$, respectively.

\subsection{Discussions}
There might be some confusion between the proposed pre-training model PROP and those weak supervision methods \cite{asadi2011pseudo, dehghani2017neural,macavaney2019content} in IR, which also leverage some classical models (e.g., BM25) to train neural ranking models. In fact, there are three major differences between the two. For those weak supervision methods: 1) Both queries and documents are available but relevance labels are missing; 2) The learning objective of weak supervision is the same as the final ranking objective; 3) The weak supervision is typically designed for each specific retrieval task. In contrary, for PROP: 1) Only documents are available while either queries or relevance labels are missing; 2) The PROP objective is not the same as the final ranking objective; 3) The pre-trained PROP model could be fine-tuned on a variety of downstream ranking tasks.

Among the pre-training objectives, the ROP objective in PROP belongs to the category of model-based pre-training objective, where the labels are produced by some automatic model rather than simple MASKs. Similar pre-training objectives in this category include Electra~\cite{clark2020electra} which leverages a generative model to replace masked tokens for pre-training the language model, and PEGASUS~\cite{zhang2019pegasus} which leverages the ROUGE1-F1 score to select top-m sentences for pre-training the abstractive summarization.

\section{Experiments}
In this section, we conduct experiments to verify the effectiveness of PROP on benchmark collections.
\subsection{Datasets}
We first introduce the two large text corpora for pre-training and five downstream ad-hoc retrieval datasets. 
\subsubsection{Pre-training Corpus}
We use two large document corpora, including English Wikipedia and MS MARCO Document Ranking dataset, to pre-train PROP since (1) They are publicly available and easy to collect; (2) A large collection of documents in these datasets could well support our pre-training method. 

\begin{itemize} [leftmargin=*]
\item \textbf{English Wikipedia} contains tens of millions of documents which has been widely used in many pre-training methods and we download the latest dump\footnote{https://dumps.wikimedia.org/} and extract the text with a public script\footnote{https://github.com/attardi/wikiextractor}.

\item \textbf{MS MARCO Document Ranking dataset} is another large-scale document collection which contains about 4 million available documents. This dataset was used in the TREC Deep Learning Track 2019\footnote{https://microsoft.github.io/TREC-2019-Deep-Learning/} and 2020\footnote{https://microsoft.github.io/TREC-2020-Deep-Learning/}. Documents are extracted from real Web documents using the Bing search engine.
\end{itemize} 

By pre-training PROP on English Wikipedia and MS MARCO Document Ranking dataset respectively, we obtain two types of models denoted as \textbf{PROP$_{\text{Wikipedia}}$} and \textbf{PROP$_{\text{MSMARCO}}$}.  

\begin{table}[t]
\renewcommand{\arraystretch}{1.6}
\setlength\tabcolsep{10pt}
  \caption{Statistics of the ad-hoc retrieval datasets}
  \label{tab:datasets}
  \begin{tabular}{ccccc}
    \toprule
    dataset& \#genre & \#queries & \#documents \\
    \midrule
    Robust04 & news & 250 & 0.5M\\ 
    ClueWeb09-B & web pages & 150 & 50M\\ 
    Gov2 & .gov pages & 150 & 25M \\ 
    MQ2007 & .gov pages & 1,692 & 25M \\ 
    MQ2008 & .gov pages & 784 & 25M \\ 
  \bottomrule
\end{tabular}
\end{table}

\subsubsection{Downstream Datasets} To verify the effectiveness of PROP, we conduct experiments on 5 representative ad-hoc retrieval datasets. 

\begin{itemize} [leftmargin=*]
\item \textbf{Robust04} consists of 250 queries and 0.5M news articles, whose topics are collected from TREC 2004 Robust Track.

\item \textbf{ClueWeb09-B} is a large Web collection with 150 queries and over 50M English documents, whose topics are accumulated from TREC Web Track 2009, 2010, and 2011. 

\item \textbf{Gov2} is a crawl of the .gov domain Web pages with 25M documents. We use 150 topic queries that are accumulated from TREC Terabyte Tracks 2004, 2005, and 2006. 

\item \textbf{Million Query Track 2007 (MQ2007)} is a LETOR~\cite{qin2013introducing} benchmark dataset with 1692 queries, which uses the Gov2 Web collection. 

\item \textbf{Million Query Track 2008 (MQ2008)} is another LETOR benchmark dataset with 784 queries, which also leverages the Gov2 Web collection.
\end{itemize} 

Note that ClueWeb09-B is filtered to the set of documents with spam scores in the $60^{th}$ percentile, using the Waterloo Fusion spam scores~\cite{cormack2011efficient}. The detailed statistics of these datasets are shown in Table~\ref{tab:datasets}.
As we can see, there is a significant difference between the number of queries and documents, which poses the challenge of training deep neural models with such a few queries.
Therefore, pre-training on large text corpus to learn universal properties can be beneficial for downstream ad-hoc retrieval tasks and avoid training deep neural model from scratch.

\subsection{Baselines}
We adopt three types of baseline methods for comparison, including traditional retrieve models, pre-trained models, and previous state-of-the-art neural ranking models.

For traditional retrieval models, we take two representative ranking methods:
\begin{itemize} [leftmargin=*]
\item \textbf{QL}: Query likelihood model ~\cite{zhai2017study} is one of the best performing language models based on Dirichlet smoothing. 
\item \textbf{BM25}: The BM25 formula ~\cite{robertson1994some} is another highly effective retrieval model that represents the classical probabilistic retrieval model. 
\end{itemize}

The pre-trained models include:
\begin{itemize} [leftmargin=*]
\item \textbf{BERT}: The key technical innovation of BERT ~\cite{devlin2018bert} is applying the multi-layer bidirectional Transformer encoder architecture for language modeling. BERT uses two different types of pre-training objectives, including Masked Language Model (MLM) and Next Sentence Prediction (NSP). Currently, BERT has become a strong baseline model for ad-hoc retrieval tasks due to its powerful contextual language representations. Different from  passage-level and sentence-level approaches~\cite{dai2019deeper,yang2019simple}, we truncate the single input sequence of the concatenated query and document to BERT's max-length limit.

\item \textbf{Transformer$_{ICT}$}: Inverse Cloze Task (ICT) \cite{lee2019latent} is specifically designed for passage retrieval in QA scenario which teaches model to predict the removed sentence given a context text. As observed in Chang et.al~\cite{chang2020pre}, ICT outperforms BFS and WLP task. Thus, we only choose ICT as the baseline for comparison.
We pre-train the Transformer model on Wikipedia corpus with ICT and MLM for a fair comparison, and other experimental settings are set the same as PROP.

\end{itemize}

Besides the above baselines, we also compare PROP with existing state-of-the-art models (SOTA) on these five datasets, including CEDR-KNRM~\cite{macavaney2019cedr} on Robust04, BERT-maxP~\cite{dai2019deeper} on ClueWeb09-B, NWT~\cite{guo2016semantic} on Gov2, and HiNT~\cite{fan2018modeling} on MQ2007 and MQ2008. We only fetch the best results of these models from the original paper. 
For ClueWeb09-B, previous SOTA BERT-maxP from Dai and Callan \cite{dai2019deeper} used a different set of queries, and thus we use their implementation to run the same query set for comparison. 

\subsection{Evaluation Methodology}
Given the limited number of queries for each collection, we conduct 5-fold cross-validation to minimize overfitting without reducing the number of learning instances. 
The parameters for each model are tuned on 4-of-5 folds. The final fold in each case is used to evaluate the optimal parameters. 
As for evaluation measures, two standard evaluation metrics, i.e., normalized discounted cumulative gain (nDCG) and precision (P), are used in experiments.    
Specifically, for Robust04, ClueWeb09-B and Gov2, we report normalized discounted cumulative gain at rank 20 (NDCG@20) and precision at rank 20 (P@20) following existing works~\cite{dai2019deeper,guo2016deep,macavaney2019cedr}. 
For MQ2007 and MQ2008, we report two official metrics used in LETOR 4.0: precision at rank 10 (P@10) and normalized discounted cumulative gain at rank 10 (NDCG@10) following existing works~\cite{pang2017deeprank,fan2018modeling}, since there are less document candidates for each dataset on the two datasets.

\subsection{Implementation Details}
Here, we describe the implementation details of PROP, including model architecture, pre-training process and fine-tuning process.

\subsubsection{Model Architecture} We use the Transformer encoder architecture similar to BERT$_{base}$ version \cite{devlin2018bert}, where the number of layers is 12, the hidden size is 768, the feed-forward layer size is 3072, the number of self-attention heads is 12, and the total parameters is 110M. 
For a fair comparison, PROP, BERT and Transformer$_{ICT}$ use the same model architecture in experiment. Specifically, we use the popular transformers library PyTorch-Transformers\footnote{https://github.com/huggingface/transformers} for the implementation of PROP.

\subsubsection{Pre-training Process} 
For representative word sets sampling, the expectation of interval $\lambda$ is set to $3$. 
To avoid sampling frequent words, we perform stopwords removal using the INQUERY stop list, discard the words that occur less than 50 times and use a subsampling of frequent words with sampling threshold of $10^{-5}$ as suggested by Word2Vec\footnote{https://github.com/tmikolov/word2vec}. Word sets are sampled with replacement, i.e. the probability for each word remains the same for multi-sampling.
We sample $5$ pairs of word sets for each document.

For the representative words prediction, we lowercase the pre-training text and do not perform stemming or stop words removal. 
The single input sequence which is concatenated by the word set and the document, is fed to PROP. We use Adam optimizer with a linear warm-up over the first 10\% steps and linear decay for later steps, and the learning rate is set to $2e-5$. Dropout with probability of 0.1 is applied on all layers. 
We train with batch size of 128 and sequence length of 512 for about 100-300k steps. 
Considering the large cost of training from scratch, we adopt the parameters of BERT$_{base}$ released by Google\footnote{https://github.com/google-research/bert} to initialize the Transformer encoder. 
We pre-train PROP on 4 Nvidia Telsa V100-32GB GPUs. 

For the masked language modeling, following BERT, we randomly select 15\% words in the input document, and the selected words are (1) 80\% of time replaced by a mask token [MASK], or (2) 10\% of time replaced by a random token, or (3) 10\% of time unchanged. Note that sampled words in representative word sets are not considered to be masked.

\subsubsection{Fine-tuning Process} 
We adopt a re-ranking strategy for efficient computation. An initial retrieval is performed using the Anserini toolkit with BM25 model to obtain the top 200 ranked documents. We then use PROP to re-rank these top candidate documents.
For all five downstream datasets, we conduct 5-fold cross-validation where each iteration uses three folds for training, one for validation, and a final held-out fold for testing. We employ a batch size among 16 and 32 and select the best fine-tuning learning rate of Adam among $1e-5$ and $2e-5$ on the validation set. 
For Robust04, ClueWeb09-B and Gov2 datasets, we perform the evaluation by using the five folds provided by Huston and Croft~\cite{huston2014parameters}. And for MQ2007 and MQ2008 datasets, we follow the data partition in LETOR4.0~\cite{qin2013introducing}. 
For all pre-trained models including BERT, Transformer$_{ICT}$ and PROP, we use raw text as the input. 
The reason is that using standard stop words removal and words stemming will hurt performance for these pre-trained models in the fine-tuning phase since it is inconsistent with pre-training process.

\begin{table*}[t]
\renewcommand{\arraystretch}{1.6}
\setlength\tabcolsep{2pt}
  \caption{Comparisons between PROP and the baselines. $\ast, \dag$ and $\ddag$ indicate statistically significance with $p-value \le 0.05$ over $\textbf{BM25}$, $\textbf{BERT}$ and Transformer$_{ICT}$, respectively.}
  \label{tab:main_res1}
  \begin{tabular}{ccccccccccccccc}
    \toprule
    \multirow{2}{*}{Model} & \multicolumn{2}{c}{Robust04} & &\multicolumn{2}{c}{ClueWeb09-B} && \multicolumn{2}{c}{Gov2}  && \multicolumn{2}{c}{MQ2007} && \multicolumn{2}{c}{MQ2008}\\
    \cline{2-3} \cline{5-6} \cline{8-9}  \cline{11-12}  \cline{14-15}  
  & nDCG@20 & P@20 & &nDCG@20 & P@20 && nDCG@20 & P@20 && nDCG@10 & P@10&& nDCG@10 & P@10  \\
    \midrule
     QL & 0.413 & 0.367 && 0.225 &0.326 & & 
     0.409 &0.510&&0.423&0.371&&0.223&0.241\\
    BM25 & 0.412 &0.363& &0.230 & 0.334& &0.421 &0.523&&0.414&0.366&&0.220&0.245\\
    Previous SOTA& \textbf{0.538} 
    &\textbf{0.467}&& 0.296& -&& 0.422 & 0.524 && 0.490&
      0.418&&
      0.244&
      0.255\\
    \midrule
      BERT& 0.459$^{\ast}$  & 0.389$^{\ast}$ && 0.295$^{\ast}$& 0.367$^{\ast}$  && 0.495$^{\ast}$& 0.586$^{\ast}$ &&0.506$^{\ast}$&0.419$^{\ast}$&&0.247$^{\ast}$&0.256$^{\ast}$ \\
      Transformer$_{ICT}$ & 0.460$^{\ast}$ & 0.388 $^{\ast}$&& 0.298$^{\ast}$ & 0.369$^{\ast}$ && 0.499$^{\ast\dag}$&0.587$^{\ast}$ && 0.508$^{\ast}$ & 0.420$^{\ast}$ && 0.245$^{\ast}$ & 0.256$^{\ast}$ \\
    \bottomrule
    \tabincell{c}{PROP$_{Wikipedia}$} & \textbf{0.502}$^{\ast\dag\ddag}$& \textbf{0.421}$^{\ast\dag\ddag}$& &
    0.316$^{\ast\dag\ddag}$&
    0.384$^{\ast\dag\ddag}$ &&
    0.519$^{\ast\dag\ddag}$ &
    0.593$^{\ast\dag\ddag}$ &&
    \textbf{0.523}$^{\ast\dag\ddag}$ &
    \textbf{0.432}$^{\ast\dag\ddag}$ &&
    0.262$^{\ast\dag\ddag}$ &
    0.267$^{\ast\dag\ddag}$ 
    \\
    \tabincell{c}{PROP$_{MS MARCO}$} &
    0.484$^{\ast\dag\ddag}$ &
    0.408$^{\ast\dag\ddag}$ &&
    \textbf{0.329}$^{\ast\dag\ddag}$ & 
    \textbf{0.391}$^{\ast\dag\ddag}$ && 
    \textbf{0.525}$^{\ast\dag\ddag}$ &
    \textbf{0.594}$^{\ast\dag\ddag}$  &&
    0.522$^{\ast\dag\ddag}$ & 
    0.430$^{\ast\dag\ddag}$ &&
    \textbf{0.266}$^{\ast\dag\ddag}$ & 
    \textbf{0.269}$^{\ast\dag\ddag}$
    \\ 
    \bottomrule
  \end{tabular}
\end{table*}

\subsection{Baseline Comparison}\label{exp}

The performance comparisons between PROP and baselines are shown in Table~\ref{tab:main_res1}. 
We can observe that: 
(1) Traditional retrieval models (i.e., \textit{QL} and \textit{BM25}) are strong baselines which perform pretty well on all downstream tasks. 
(2) By automatically learning text representations and relevance matching patterns between queries and documents, previous state-of-the-art neural ranking models can achieve better results than traditional retrieval models. For Robust04 and ClueWeb09-B, BERT-based models, i.e. CEDR and BERT-maxP, achieve significant improvements over \textit{QL} and \textit{BM25}, while traditional neural ranking models including DRMM and NWT shows slight improvements over \textit{QL} and \textit{BM25} on other three datasets (i.e., Gov2, MQ2007, and MQ2008).   
One possible reason is that it is difficult for a deep neural model training from scratch with such a few supervised pairs.
(3) The improvements of \textit{BERT} and \textit{Transformer$_{ICT}$} over previous \textit{SOTA} on Gov2, MQ2007 and MQ2008 datasets, demonstrate that pre-training and
fine-tuning are helpful for downstream tasks. 

When we look at the two types of \textit{PROP} models pre-trained on Wikipedia and MS MARCO respectively, we find that: \textit{PROP$_{Wikipedia}$} achieves better results than \textit{PROP$_{MSMARCO}$} on Robust04, while \textit{PROP$_{MSMARCO}$} performs better than \textit{PROP$_{Wikipedia}$} on ClueWeb09-B.
The reason might be that the news articles in Robust04 are similar with the well-formed articles in Wikipedia while the Web pages in ClueWeb09-B are similar with the Web documents in MS MARCO. 
These results suggest that employing the pre-trained model from a related domain for the downstream task is much more effective.

Finally, we can see that the best \textit{PROP} model achieves the best performance in terms of all the evaluation metrics in 4 of 5 datasets. 
The observations are as follows: 
(1) PROP outperforms traditional retrieval models (i.e., \textit{QL} and \textit{BM25}) by a substantial margin. For example, the relative improvement of \textit{PROP} over \textit{BM25} is about 46\% in terms of nDCG@20 on ClueWeb09-B. 
The results indicate the effectiveness of our pre-training method for ad-hoc retrieval. 
(2) Compared with previous \textit{SOTA}, the relative improvements are about 8.9\%, 24.4\%, 6.7\% and 9\% in terms of nDCG@20 for ClueWeb09-B, Gov2, MQ2007 and MQ2008 respectively. 
For Robust04, CEDR-KNRM is better than PROP since CEDR integrates BERT's representations into existing neural ranking models, e.g. KNRM.
Nevertheless, the results demonstrate that pre-training on a large corpus and then fine-tuning on downstream tasks is better than training a neural deep ranking model from scratch. 
(3) The improvements of PROP over existing pre-trained models (i.e., \textit{BERT} and \textit{Transformer$_{ICT}$}) indicate that designing a pre-training objective tailored for IR with a good theoretical foundation is better than directly applying pre-training objectives from NLP on IR tasks. 

\begin{table}[t]
\renewcommand{\arraystretch}{1.6}
\setlength\tabcolsep{1.pt}
  \caption{Impact of pre-training objectives. $\dag$ indicates statistically significance with $p-value < 0.05$.}
  \label{tab:obj}
  \begin{tabular}{ccccccc}
    \toprule
       \multirow{2}{*}{} & \multicolumn{3}{c}{nDCG@20} &&\multicolumn{2}{c}{nDCG@10}\\
    \cline{2-4} \cline{6-7}
    & Robust04 & ClueWeb09-B & Gov2 && MQ2007 & MQ2008
    \\
    \midrule
    w/ MLM  & 0.467  & 0.306 & 0.503 && 0.511 & 0.249  \\
    w/ ROP  & 0.481$^{\dag}$ & 0.321$^{\dag}$ & 0.519$^{\dag}$ && 0.520$^{\dag}$ & 0.262$^{\dag}$ \\
   w/ ROP+MLM& \textbf{0.484$^{\dag}$} & \textbf{0.329$^{\dag}$}  & \textbf{0.525$^{\dag}$} & & \textbf{0.522$^{\dag}$} & \textbf{0.266$^{\dag}$}\\
  \bottomrule
\end{tabular}
\end{table}

\subsection{Impact of Pre-training Objectives} 
In this section, we investigate the effect of different  pre-training objectives in PROP.  
Specifically, we pre-train the Transformer model with the ROP and MLM objective respectively on MS MARCO under the same experiment settings in PROP. 
As shown in Table~\ref{tab:obj}, we report the nDCG results on 5 downstream tasks. 
We can see that: 
(1) Pre-training with MLM on MS MARCO shows slight improvements over BERT pre-trained on BooksCorpus and English Wikipedia (as shown in Table~\ref{tab:main_res1}). 
It indicates that good representations obtained by MLM may not be sufficient for ad-hoc retrieval tasks. 
(2) Pre-training with ROP achieves significant improvements over MLM on all downstream tasks, showing the effectiveness of ROP tailored for IR. 
(3) By pre-training jointly with the ROP and MLM objective, PROP achieves the best performance on all downstream tasks. 
It indicates that the MLM objective which brings good contextual representations and the ROP objective which resembles the relevance relationship for ad-hoc retrieval tasks can contribute together. 

\begin{table*}[t]
\renewcommand{\arraystretch}{1.6}
\setlength\tabcolsep{5pt}
  \caption{Impact of Further Pre-training on Target Tasks. Two-tailed t-tests demonstrate the improvements of PROP with further pre-training to that without further pre-training are statistically significant ($\dag$ indicates $\text{p-value} < 0.05$).}
  \label{tab:fp}
  \begin{tabular}{cccccccc}
    \toprule
    \multirow{2}{*}{} & \multicolumn{3}{c}{nDCG@20} &&\multicolumn{2}{c}{nDCG@10}\\
    \cline{2-4} \cline{6-7}
    & Robust04 & ClueWeb09-B & Gov2 && MQ2007 & MQ2008
    \\
    \midrule
    Without Further Pre-training & 0.502 & 0.329 & 0.525 && 0.523 & 0.266\\
    Further Pre-training & 0.506$^{\dag}$ & 0.334$^{\dag}$ & 0.531$^{\dag}$ &&0.526$^{\dag}$&0.270$^{\dag}$ \\

  \bottomrule
\end{tabular}
\end{table*}

\begin{table}[t]
\renewcommand{\arraystretch}{1.5}
\setlength\tabcolsep{1.5pt}
  \caption{Impact of Different Sampling Strategies. Two-tailed t-tests demonstrate the improvements of document language model-based sampling to the random sampling strategy are statistically significant ($\dag$ indicates $\text{p-value} < 0.05$). }
  \label{tab:ss}
  \begin{tabular}{ccccccc}
    \toprule
   \multirow{2}{*}{} & \multicolumn{3}{c}{nDCG@20} &&\multicolumn{2}{c}{nDCG@10}\\
    \cline{2-4} \cline{6-7}
    & Robust04 & ClueWeb09-B & Gov2 && MQ2007 & MQ2008
    \\
    \midrule
    Random  & 0.471  &0.304 & 0.505 && 0.513 & 0.252  \\
    docLM-based  & 0.493$^{\dag}$ & 0.317$^{\dag}$ & 0.517$^{\dag}$ && 0.516$^{\dag}$ & 0.257$^{\dag}$ \\
  \bottomrule
\end{tabular}
\end{table}

\subsection{Impact of Sampling Strategies}

\begin{figure}[t]
	\centering
		\includegraphics[scale=0.28]{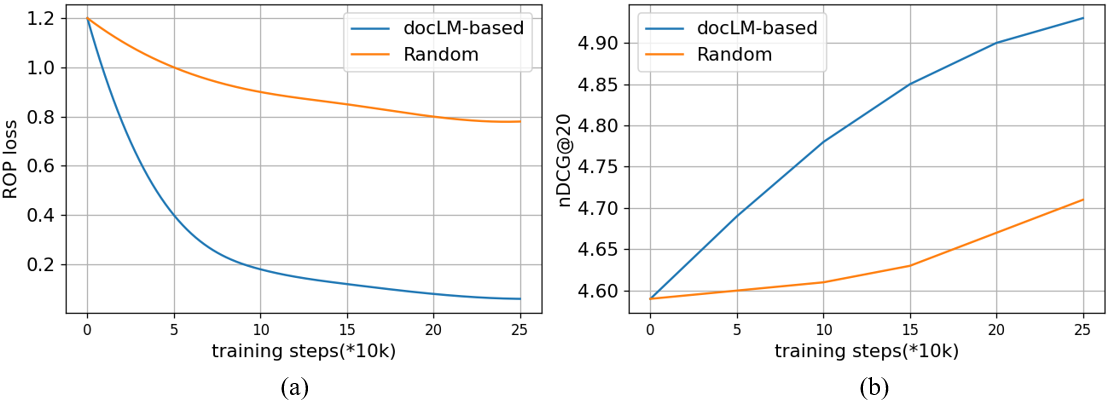}
		  \caption{(a) ROP learning curve on Wikipedia over the pre-training steps. (b) The test performance curve on Robust04 in terms of nDCG@20 over the pre-training steps.}
  \label{fig:ss}
\end{figure}

As described in Section~\ref{sec:pre-method}, the representative word sets are sampled according to the document language model.  
To investigate the impact of different sampling strategies, we compare the document language model-based sampling strategy  (docLM-based sampling for short) with the random sampling strategy, which randomly samples words from the corpus vocabulary independently. 
Specifically, we pre-train the Transformer model on English Wikipedia only with the ROP objective using docLM-based sampling and random sampling respectively. 
The loss curve of the ROP objective over the pre-training steps  using different sampling strategies is depicted in Figure~\ref{fig:ss} (a). 
We can find that PROP based on the docLM-based sampling strategy converges much faster than that based on the random sampling strategy.

Moreover, we pre-train the Transformer model at most 250K steps for both sampling strategies, and further fine-tune them on the five downstream datasets. 
As shown in Table~\ref{tab:ss}, we report the nDCG@20 results on Robust04, ClueWeb09-B, and Gov2 datasets, and the nDCG@10 results on MQ2007 and MQ2008 datasets. 
We find that PROP based on the docLM-based sampling strategy can achieve significantly better results than that based on the random sampling strategy. 
We also show the test performance curve in terms of nDCG@20 on Robust04 over the pre-training steps in Figure~\ref{fig:ss}(b). We can observe that PROP based on the docLM-based sampling strategy improves the performance much faster than that based on the random sampling strategy. 

All the above results indicate that the docLM-based sampling strategy is a more suitable way than the random sampling strategy to generate representative word sets for a document.
The reason might be that the document language model roots in a good theoretical IR foundation and thus contributes to the efficiency and effectiveness of the pre-training process. 

\subsection{Further Pre-training on Target Tasks}

Here, we analyze the impact of further pre-training on the  document collections in the target tasks to see how much performance could be improved.   
Specifically, we further pre-train PROP$_{Wikipedia}$ on the document collections of Robust04 and MQ2007 respectively, and  PROP$_{MSMARCO}$ on the document collections of ClueWeb09-B, Gov2 and MQ2008 respectively. 
As shown in Table~\ref{tab:fp}, we can see that PROP with further pre-training on the document collection in the target tasks outperforms that without further pre-training. 
The results demonstrate that further pre-training on the related-domain corpus could improve the ability of PROP and achieve better performance on the downstream tasks, which is  quite consistent with the previous findings \cite{howard2018universal}.  
However, the improvement of further pre-training over without further pre-training on MQ2007 dataset is less than that on other four datasets. 
The reason might be that MQ2007 has much more queries than other datasets, and enough in-domain information can be well captured  during the fine-tuning process.

\begin{figure*}[t]
	\centering
		\includegraphics[scale=0.5]{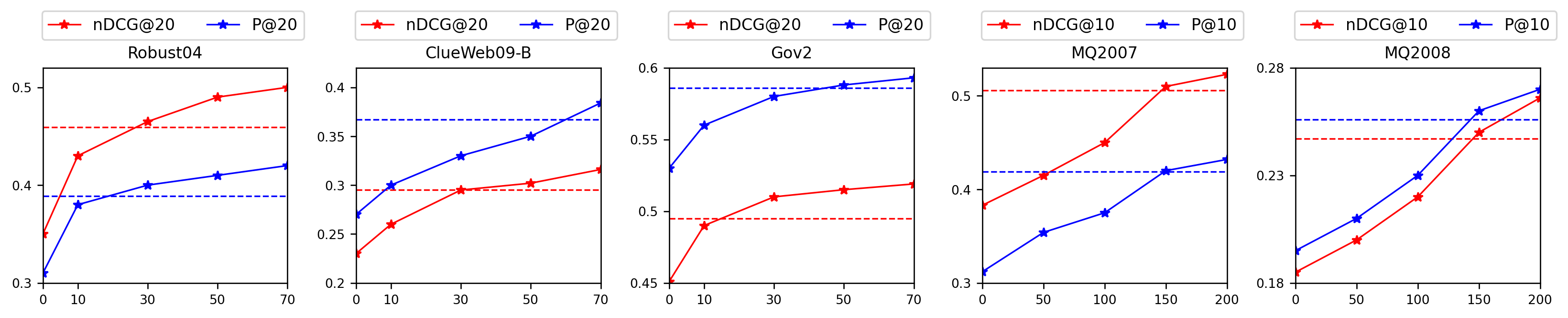}
		  \caption{Fine-tuning with limited supervised data. The solid lines are PROP fine-tuned using 0 (zero shot), 10, 30, 50, and 70 queries for Robust04, ClueWeb09-B and  Gov2 datasets, using 0 (zero shot), 50, 100, 150, and 200 queries for MQ2007 and MQ2008 datasets. The dashed lines are BERT fine-tuned using the full queries.}
  \label{fig:fs}
\end{figure*}

\subsection{Zero- and Low-Resource Settings}
In real-world practice, it is often time-consuming and difficult to collect a large number of relevance labels in IR evaluation.  
To simulate the low-resource IR setting, we pick the first 10, 30, 50, 70 queries from Robust04, ClueWeb09-B, and Gov2, and the first 50, 100, 150, 200 queries from MQ2007 and MQ2008 to fine-tune PROP$_{Wikipedia}$. 
We fine-tune the models with batch size as three different values (i.e., 4, 8, 16), learning rate as two different values (i.e., 1e-5, 2e-5), and pick the checkpoint with the best validation performance. 
As shown in Figure~\ref{fig:fs}, we can find that:  
(1) \textit{PROP} fine-tuned on limited supervised data can achieve comparable performance with BERT fine-tuned on the full supervised datasets in terms of nDCG and Precision. 
For example, PROP fine-tuned with only 30 queries has outperformed BERT on Robust04, ClueWeb09-B, and Gov2 datasets.   
(3) Furthermore, fine-tuning \textit{PROP} with only 10 queries can achieve comparable results with traditional retrieval models (i.e., \textit{QL}and \textit{BM25}) for Robust04, ClueWeb09-B, and Gov2. 
The results demonstrate that by fine-tuning with small numbers of supervised pairs, PROP is able to adapt to the target task quickly. 
(4) Under the zero resource setting, for example, PROP can achieve about 90\% performance of BERT fine-tuned on the full Gov2 dataset in terms of nDCG@20.

\section{RELATED WORK}
In this section, we briefly review two lines of related work, i.e., pre-trained language models and pre-training objectives for IR.

\subsection{Pre-trained Language Models}
Recently, pre-trained language representation models such as OpenAI GPT~\cite{radford2018improving}, BERT~\cite{devlin2018bert} and XLNET~\cite{yang2019xlnet}, have led to significant improvements on many NLP tasks. 
The key idea is that firstly pre-training a large neural architecture on massive amount of unlabeled data and then fine-tuning on downstream tasks with limited supervised data. Transformer~\cite{vaswani2017attention} has become the mainstream architecture of pre-trained models due to its powerful capacity. 
Pretext tasks, such as probabilistic language modeling~\cite{bengio2003neural}, Masked language modeling~\cite{taylor1953cloze, yang2019xlnet} and Permuted Language Modeling (PLM)~\cite{yang2019xlnet}, have been proved effective in NLP since they can learn universal language representations and contribute to the downstream NLP tasks.

BERT, as the most prominent one among existing pre-trained models, pre-training the Transformer with MLM and NSP, to obtain contextual language representations and sentence-pair representations. 
Directly applying BERT to IR can achieve new state-of-the-art performance. 
A simple approach is to feed the query–document pair to BERT and use an MLP or more complicated module on top of BERT’s  output to produce relevance score. 
Nogueira et al.~\cite{nogueira2019passage,nogueira2019multi} trained BERT model on MS MARCO passage ranking task and TREC CAR using pointwise and pairwise approaches. 
Dai et al.~\cite{dai2019deeper} and Yang et al.~\cite{yang2019simple} adopted passage-level and sentence-level methods for addressing the document length issue respectively, i.e. applying inference on sentences/passages individually, and then aggregating sentence/passage scores to produce document scores. 
MacAvaney et al.~\cite{macavaney2019cedr} integrated the representation of BERT’s [CLS] token into existing neural ranking models such as DRMM~\cite{guo2016deep}, PACRR~\cite{hui2017pacrr} and KNRM~\cite{xiong2017end}. 
Despite the success BERT has achieved in IR community,  designing pre-training objectives specially for IR is still of great potential and importance.


\subsection{Pre-training Objectives for IR}
 There has been little effort to design pre-training objectives towards ad-hoc retrieval. Most related work in this direction focused on passage retrieval in question answering (QA)~\cite{lee2019latent,chang2020pre}.
For example, Lee et al. ~\cite{lee2019latent} proposed Inverse Cloze Task (ICT) for passage retrieval, which randomly samples a sentence from passage as pseudo query and takes the rest sentences as the document. However, this method may lose the important exact matching patterns since the pseudo query is removed from the original document.
In \cite{chang2020pre}, Chang et al.~ proposed another two additional pre-training objectives, i.e., Body First Selection (BFS) and Wiki Link Prediction(WLP). BFS randomly samples a sentence in the first section of a Wikipedia page as pseudo query and the document is a randomly sampled paragraph from the same page. 
WLP chooses a random sentence in the first section of a Wikipedia page as pseudo query, then a document is sampled from another page where there is a hyperlink between these two pages. 
However, such pre-training objectives rely on special structure of the document (e.g., multiple paragraph segmentations and hyperlinks), which hinder the method to be applied on general text corpus.
In summary, all these above pre-triaining tasks attempt to resemble the relevance relationship between natural language questions and answer passages. As demonstrated in our experiments, when applying pre-trained models based on these tasks to ad-hoc retrieval, marginal benefit could be observed on typical benchmark datasets.

\section{CONCLUSION}
In this paper, we have proposed PROP, a new pre-training method tailed for ad-hoc retrieval. The key idea is to pre-train the Transformer model to predict the pairwise preference between the two sets of words given a document, jointly with the MLM objective. PROP just needs to pre-train one model and then fine tune on a variety of downstream ad-hoc retrieval tasks. Through experiments on 5 benchmark ad-hoc retrieval datasets, PROP achieved significant improvements over the baseline without pre-training or with other pre-training methods. We also show that PROP can achieve strong performance under both the zero- and low-resource IR settings.

For future work, we would like to go beyond the ad-hoc retrieval, and try to test the ability of PROP over other types of downstream IR tasks, such as passage retrieval in QA or response retrieval in dialog systems. We will also investigate new ways to further enhance the pre-training tailored for IR.


\section{Acknowledgments}

This work was supported by Beijing Academy of Artificial Intelligence (BAAI) under Grants No. BAAI2019ZD0306, and funded by the National Natural Science Foundation of China (NSFC) under Grants No. 61722211, 61773362, 61872338, 62006218, and 61902381, the Youth Innovation Promotion Association CAS under Grants No. 20144310, and 2016102, the National Key RD Program of China under Grants No. 2016QY02D0405, the Lenovo-CAS Joint Lab Youth Scientist Project, the K.C.Wong Education Foundation, and the Foundation and Frontier Research Key Program of Chongqing Science and Technology Commission (No. cstc2017jcyjBX0059).

\bibliographystyle{ACM-Reference-Format}
\bibliography{sample-base}


\end{document}